\documentclass[twocolumn]{aastex62}
\usepackage{graphicx}
\usepackage[suffix=]{epstopdf}
\usepackage{natbib}
\usepackage{amsmath}
\usepackage{xspace}

\newcommand{\Kepler}{\textsl{Kepler}\xspace}
\newcommand{\TESS}{\textsl{TESS}\xspace}

\begin{document}
\title{SETI in the Spatio-Temporal Survey Domain
}

\shorttitle{Survey SETI}
\shortauthors{Davenport}

\author{James. R. A. Davenport}
\affiliation{Department of Astronomy, University of Washington, Seattle, WA 98195, USA}
\altaffiliation{DIRAC Fellow}

\begin{abstract}
Traditional searches for extraterrestrial intelligence (SETI) or ``technosignatures'' focus on dedicated observations of single stars or regions in the sky to detect excess or transient emission from intelligent sources. The newest generation of synoptic time domain surveys enable an entirely new approach: spatio--temporal SETI, where technosignatures may be discovered from spatially resolved sources or multiple stars over time. 
Current optical time domain surveys such as ZTF and the Evryscope can probe 10--100 times more of the ``Cosmic Haystack'' parameter space volume than many radio SETI investigations. Small-aperture, high cadence surveys like Evryscope can be comparable in their Haystack volume completeness to deeper surveys including LSST. Investigations with these surveys can also be conducted at a fraction of the cost of dedicated SETI surveys, since they make use of data already being gathered. 
However, SETI methodology has not widely utilized such surveys, and the field is in need of new search algorithms that can account for signals in both the spatial and temporal domains.
Here I describe the broad potential for modern wide-field time domain optical surveys to revolutionize our search for technosignatures, and illustrate some example SETI approaches using transiting exoplanets to form a distributed beacon.
\end{abstract}

\section{Introduction}

Despite more than half a century of activity, the Search for Extra-Terrestrial Intelligence (SETI) -- also described as the search for ``technosignatures'' -- is a field of study very much in its infancy. 
The lack of SETI research by the astronomical community is partially due to funding limitations in recent decades, as well as social pressures against studying SETI experienced by professional astronomers and organizations \citep{wright2018b}. 
Indeed, the volume of parameter space for SETI (or the search for a needle amongst the ``Cosmic Haystack'') at radio wavelengths has been barely explored \citet{wright2018c}, and thus many obvious signals may be awaiting our discovery. 
However, the new era of surveys and data-driven discovery in astronomy holds great promise for revolutionizing SETI.

Traditional SETI work requires time-consuming observations, often through dedicated monitoring of nearby stars by radio telescopes. Such observing campaigns are expensive and difficult to obtain given the competitive nature of telescope allocation. Recent progress for systematic ``technosignature'' searches has been made by Breakthrough Listen \citep{worden2017,isaacson2017}, primarily at radio wavelengths \citep[e.g.][]{price2018}.

The SETI conundrum has always been not only ``{\it What should we look for?}'', but  also {\it ``When, where, and how}''. 
While \citet{wright2018c} suggest the completeness of our search of the ``Cosmic Haystack'' at radio wavelengths is akin to the ratio of a small swimming pool compared to the Earth's oceans ($\sim$10$^{-20}$), SETI at optical wavelengths is surely many orders of magnitude less complete.  
Current optical-SETI is typically focused on searches for nanosecond pulses with specialized wide-field high-speed imaging \citep{howard2004}, or unusual spectroscopic features \citep{isaacson2019}.
Wide-field time domain photometric surveys enable us to greatly expand our search in two key dimensions of parameter space: 1) collectively observing nearly the entire night sky, and 2) providing precision time-resolved monitoring over many years.

\citet{djorgovski2000} highlighted the opportunity for a new age of generalized SETI using a virtual observatory made of digital sky surveys.
New missions like the currently running Zwicky Transient Facility \citep[ZTF;][]{bellm2014} and the upcoming Large Synoptic Survey Telescope \citep[LSST;][]{lsst} will be transformative in enabling both reproducible and cost-efficient SETI. By developing search strategies that utilize public survey databases and real-time ``alert streams'', both of which are being developed to facilitate a wide range of science goals from these facilities, optical survey SETI can be conducted automatically. As new search algorithms are developed, these databases can be re-analyzed, providing a much-needed level of reproducibility and transparency that will help reduce the ``giggle factor'' stigma for SETI \citep{wright2018b}.

In this paper I explore the potential for wide-field time domain surveys in conducting a new generation of technosignature searches, starting with an overview of the motivation for doing SETI with optical surveys in \S\ref{sec:method}.
In \S\ref{sec:strategy} I describe the general principles behind technosignature searches in time domain surveys, including spatial over-densities of light curve signals and coordinated signals between multiple sources, for example.
In \S\ref{sec:haystack} I introduce several relevant surveys, and use the ``Cosmic Haystack'' search volume metric developed by \citet{wright2018c} to quantitatively compare these surveys to current radio SETI approaches. 
The data-driven spatio-temporal domain explored by these surveys necessitates new SETI approaches be developed, and in \S\ref{sec:transit} I outline an example signal where transiting planets around many stars could act as a spatially distributed beacon.
Finally in \S\ref{sec:discussion} I discuss ideas for other search approaches, and the advantages in cost and reproducibility of conducting SETI with current and upcoming optical time domain surveys.

\section{Why Optical Survey SETI?}
\label{sec:method}

Wide-field imaging surveys, and their associated infrastructure, are an ideal data source for carrying out technosignature searches that augment efforts in other wavelengths or with different observing strategies. By covering large portions of the night sky with repeated imaging, and with surveys that span years to decades in duration, we can substantially reduce the challenge of deciding {\it when} and {\it where} to look. Increasingly, multiple surveys obtain imaging that overlaps in both time and sky coverage, but with complementary properties (e.g. imaging depth, cadence, wavelength coverage), which allows for rapid vetting and characterization of rare or low signal-to-noise events. Surveys also produce important baseline information about the known objects in our Galaxy (e.g. their positions, temperatures, etc), which allows us to eliminate uncertainty on {\it what} objects being targeted for SETI.

Optical surveys are also an excellent platform for exploring {\it how} to conduct technosignature searches. The large archives of data produced from these surveys become valuable legacies that enable science long after data acquisition has concluded. Searches for new phenomena in time-domain astronomy can be carried out both in the future with new observations and in the past by re-analyzing survey archives. 
Using survey archives for SETI as opposed to targeted observations of individual stars allows us to compare the results of competing algorithms or strategies, and enables reproducible science. Search algorithms are also portable, and ideally can be used to carry out SETI on any new survey or large ensemble of observations.
As future survey data is gathered, new classes of technosignature signals may be detectable that rely on decades worth of data. 
Since any technosignature detection candidate should be the subject of great scrutiny, making the data and algorithms open and reproducible will be critical for comparison and independent validation.

Large surveys, particularly ground-based efforts like LSST, are a highly prioritized component of the current and future development landscape in astronomy \citep{astro2010}. As such, great efforts are being made to develop survey tools and technologies to enable science from these missions, such as new database technologies \citep{juric2012,zecevic2019}, real-time event broadcasting \citep{patterson2019}, and real-time data analysis frameworks \citep[e.g.][]{schwamb2019}.
The infrastructure needed to conduct technosignature searches with these large optical surveys is therefore already being developed. As with many areas of astrophysics in the ``survey era'', the biggest bottleneck for SETI with these surveys is the support (both financial and social) for researchers to develop algorithms and mine the available data.

\section{Potential Strategies}
\label{sec:strategy}

SETI with large surveys falls into two basic approaches: 1) classify every object or phenomena in the data, and any remaining outliers are technosignature candidates \citep{djorgovski2000}, or 2) search for specific ``impossible'' or unusual types of signals in the data, which I advocate for in this work. While the latter strategy does not guarantee a complete search for all SETI signals in the data, it is technically feasible for almost any time domain survey. 

Several types of signals have been suggested that may be viable for use with optical surveys. For example, \citet{villarroel2016} searched two wide field, single-epoch optical surveys for objects ``disappearing''  over decades timescales. \citet{lacki2019} outline a method to search for specular reflections of interplanetary objects, particularly in wide field surveys. \citet{arnold2005} describe the the observable signature of transits from artificial objects, which could be searched for with exoplanet transit surveys. \citet{kipping2016} find that exoplanet transits could be hidden or ``cloaked'' to an observer by using directed laser emission.

Here I describe a few general classes of technosignature signals that may be well suited to developing new detection algorithms for spatio-temporal surveys:

\begin{enumerate}
\item {\bf Unusual variability profiles or statistical distributions of fluxes }-- either on short timescales such as Boyajian's Star \citep{boyajian2015}, or long timescales such as disappearing stars \citep{villarroel2016} or occasionally missing transits \citep{kipping2016}
\item {\bf Spatial correlations of events or phenomena }-- e.g. coordination of transiting systems, or rebroadcasting events such as novae along the ``SETI ellipse''  that defines the ellipsoid for receiving synchronized signals from multiple transmitters 
\citep{makovetskii1977,lemarchand1994,tarter2001,shostak2004}
\item {\bf Spatial over-density or distribution} -- e.g. an over-density of a given phenomena, such as within the ``Earth Transit Zone'' band of stars that would see Earth as a transiting exoplanet \citep{heller2016}, or in spatial clusters
\item {\bf Unnatural patterns} -- particularly patterns of otherwise normal astrophysical variability (e.g. flares, pulsations, transits, etc), such as broadcasting a prime or fibonacci sequence using transits \citep{arnold2005,wright2016}.
\end{enumerate}

\section{Comparing Optical Surveys\\ to Radio Searches}
\label{sec:haystack}

While the potential for optical time domain surveys to advance technosignature searches has been qualitatively laid out in \S\ref{sec:method}, it is important to quantitatively compare them to traditional searches in the radio. 
\citet{wright2018c} have provided a framework to quantify the volume of observable parameter space covered by various SETI projects -- dubbed the ``Cosmic Haystack''. Though this metric was developed with radio surveys in mind, it is broadly applicable to surveys at other wavelengths that are not explicitly designed for SETI \citep[e.g. see][]{forgan2019}.

For typical optical surveys using broadband photometry \citep{djorgovski2013}, the key parameters that determine the Haystack volume coverage are the sky coverage ($\Omega$ in sq deg), total integration time per target ($T$ in sec), and survey sensitivity or per-epoch depth ($S$ in Jy).
Since these terms all act as scalers in the Haystack volume integration, surveys with different properties (e.g. few deep exposures versus many shallow exposures) can have comparably large Haystack coverage, filling out very different portions of the search volume. For maximum impact, SETI should therefore be conducted on many types of surveys.

\begin{figure}[]
\centering
\includegraphics[width=3.5in]{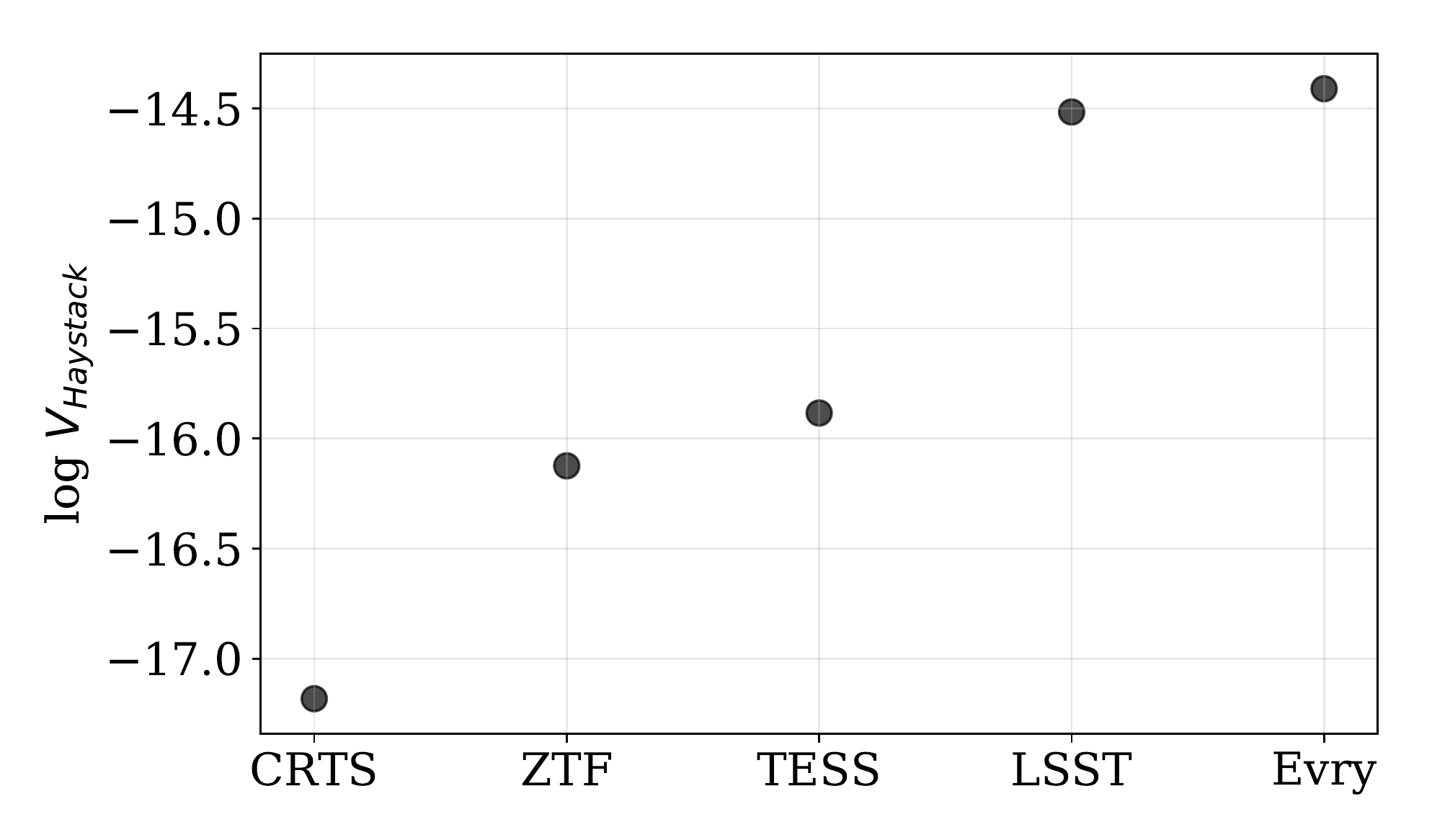}
\caption{
Comparison of the 8-D ``Cosmic Haystack'' SETI search volume fraction, defined by \citet{wright2018c}, computed using five optical surveys with varying designs. The Haystack fraction covered by these optical surveys is 1-2 dex larger than typical SETI programs conducted in the radio.
Evryscope \citet{law2015} is the best survey considered here for SETI work, narrowly beating LSST due to its wide field of view and very dense light curves.}
\label{fig:hay}
\end{figure}

As a demonstration, I have computed the Haystack volume coverage for five representative current and future optical time domain surveys, as shown in Figure \ref{fig:hay}. This includes the Catalina Real Time Transient Survey \citep[CRTS;][]{catalina}, Zwicky Transient Facility \citep[ZTF;][]{bellm2014}, the Transiting Exoplanet Survey Satellite \citep[\TESS;][]{tess}, the Large Synoptic Survey Telescope \citep[LSST;][]{lsst}, and the Evryscope \citep{law2015}. These include space-based and ground-based surveys, and vary greatly in telescope collecting area, total numbers of visits, and survey footprint design. 
I have assumed sensitivity ($S$) is the published per-visit photometric depth for each survey. The total integration time per target ($T$) was computed as the average number of visits per target in the survey multiplied by the per-visit exposure time. The same power of transmitter is assumed as in \citet{wright2018c}. 
The code to generate the figures in this manuscript is online.\footnote{\url{https://github.com/TheAstroFactory/survey_seti}}

Typical values for the Haystack fraction for radio SETI programs as presented in \citet{wright2018c}  are in the range of $\log V_{Haystack} =  -20$ to $-17$. However, these programs are mostly targeted observations of nearby stars. The largest program featured in \citet{wright2018c}, by \citet{tingay2018}, had $\Omega=400$ sq deg and $T=10800$ seconds, and yielded $\log V_{Haystack} =  -17.36$.

The wide field optical surveys I have examined here all out perform the radio programs explored by \citet{wright2018c} in their Haystack fraction. Due to its very wide field coverage, and near continuous monitoring each night, Evryscope (abbreviated as ``Evry'' in Figure \ref{fig:hay}) ranks as the best optical survey for technosignature searches of those surveyed here. The remarkable similarity in Haystack fraction between Evryscope, which uses 7-cm diameter telescopes, and the 8.4-m LSST survey reinforces the need to sample the Haystack volume with as many different survey ``shapes'' as possible. Note that since the area coverage, photometric depth, and cadence of these two surveys are so different, they will each be sensitive to very different specific SETI detection properties, such as signal repetition timescales or amplitudes.

\citet{wright2018c} likened their Haystack fraction to searching for fish in a hot tub or small swimming pool compared to the Earth's oceans. Our results for SETI with optical surveys are closer to 1-2 Olympic sized swimming pools (each 2500 m$^3$, according FINA standards) compared to the Earth's oceans\footnote{Calculation done via Wolfram$|$Alpha}. While optical surveys still only cover a tiny fraction of the total volume of parameter space, it is a substantial improvement.

\section{Example: Distribution of Transiting Exoplanets in the {\it Kepler} Field}
\label{sec:transit}

To illustrate the potential for possible technosignature searches in the spatio-temporal survey domain, in this section I demonstrate two examples of signals that rely on samples of transiting exoplanets. Transits are a promising target for SETI since they are relatively efficient to find over wide areas with surveys like \Kepler \citep{borucki2010} or \TESS, and they represent a natural object to attract attention with, known as a ``Schelling point'' \citep{wright2017}. Creating such a beacon requires either generating a transiting system with unusual properties or patterns \citep[e.g.][]{arnold2005,kipping2016,forgan2017}, or as outlined in \S\ref{sec:strategy} an unusual {\it distribution} of transiting systems. The two examples outlined here explore the latter scenario, though I note the former is also well suited to automated approaches that would make use of future surveys and should be developed further. While I do not advocate these {\it specific} metrics are ideal for SETI, they demonstrate the type of signals we should design software systems to find in modern surveys.

\subsection{Two-Point Correlation Function}
\label{sec:2pt}

The easiest conceptual method to search for a technosignature beacon in the spatial distribution of exoplanets would be looking for over-densities, i.e. regions with an unusually large number of transiting systems relative to the local density of stars. However this requires a large and very uniform sampling of exoplanets per square degree to determine if the density is unusual. Such spatially uniform samples are difficult to achieve since e.g. bright stars can saturate large regions.

Another simple statistical test is to measure the distribution of exoplanet systems using the two-point angular correlation function \citep{landy1993}. This method measures the probability of finding two sources separated by a given radial distance, and is often used in cosmology to characterize the distribution of galaxies on cosmological scales,. The two-point correlation function is easily measured thanks to generalized functions from packages such as {\tt astroML} \citep{astroMLText}, and  more efficient approaches have been produced that can scale to very large datasets \citep[e.g.][]{wang2013}.

\begin{figure}[!t]
\centering
\includegraphics[width=3in]{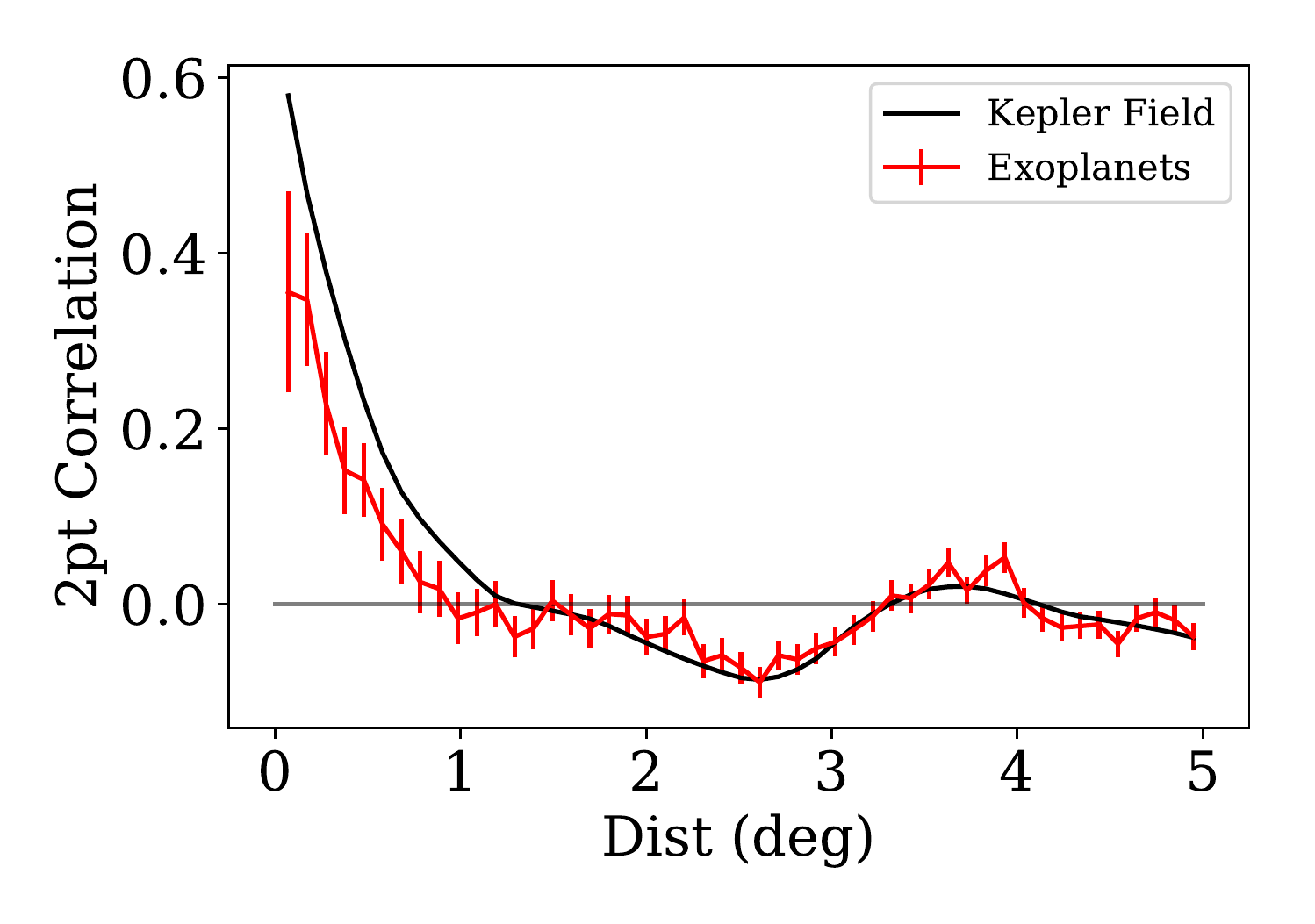}
\caption{two point correlation function of transiting exoplanets versus all stars in the kepler field. No obvious signs of engineering of the exoplanet population as a whole. exoplanets are actually slightly less correlated on small size scales than stars in general, which may be an artifact of detected planets being around brighter stars in the sample (i.e. having lower total density). 
While overly simple, this example demonstrates one possible test that could be done using existing and future exoplanet surveys.}
\label{fig:2pt}
\end{figure}

In Figure \ref{fig:2pt} I show the two-point correlation function up to spatial scales of 5 degrees for both the sample of 2,324 transiting exoplanets discovered by the \Kepler primary mission,\footnote{Catalog from \url{http://exoplanetarchive.ipac.caltech.edu}} and for 201,312 stars in the \Kepler field. Since the sample of exoplanets was relatively small, uncertainties shown in Figure \ref{fig:2pt} were estimated by bootstrapping the correlation function 500 times. Here the correlation was done using only the (RA, Dec) coordinates for each star or exoplanet system, but can also be done for additional dimensions (e.g. including distance, orbital properties, etc).

The correlation functions shown in Figure \ref{fig:2pt} are not substantively different between the exoplanet and field star populations, indicating no signs of a large-scale ``beacon'' based on the distribution of exoplanets in the \Kepler field. The slight differences in the correlation functions are not unexpected, given how sensitive this metric is to sample selection biases and non-uniformity. Indeed, even in ``simple'' scenarios of galaxy population analyses from well-studied surveys, considerable effort must be made to compare the correlation functions between different samples \citep[e.g.][]{wang2013}. As the 2-D correlation requires very homogeneous and large samples of sources to plausibly detect outliers with, it would be an inefficient metric for a civilization to base a beacon around.

\subsection{Clusters of Similarity}
\label{sec:sim}

Rather than searching for an exoplanet beacon consisting of outliers in spatial density or distribution (e.g. \S\ref{sec:2pt}), we can consider groups of systems that are unusually similar in some way. In the simplest case this could take the form of many closely clustered stars with transiting exoplanets, where each planet had the same radius, transit depth, transit ephemeris (or transit midpoint), or orbital period, for example. The assumption here is that a civilization would have engineered this unlikely set of transiting systems, like a lighthouse, in order to be noticed by early astronomers who are characterizing the exoplanet demographics of our Galaxy (i.e. us). This SETI approach relies on both the spatial and temporal information contained in exoplanet-hunting surveys like \Kepler or \TESS.
Here I demonstrate a simple algorithm for searching for such clusters of exoplanets with unusually similar orbital periods.

As with many proposed technosignature signals, the key to this spatially distributed beacon system being effective is that the similarity metric for the cluster of stars must be substantially outside the realm of possible values that nature can produce. A civilization constructing such a beacon would also need to be aware of the natural distribution of exoplanet periods in order to determine the precision needed to place their transiting bodies. Thus we don't have to arbitrarily choose which orbital periods are interesting to search over or how close in period they need to be to count as a technosignature candidate, but instead we can adopt a data-driven approach where our thresholds are empirically set based on a probability threshold. 

For this demonstration I selected the same sample used in \S\ref{sec:2pt} of 2,324 known transiting exoplanet systems within the \Kepler field, with orbital periods ranging from 0.35 days to 1,322.5 days. 
The NearestNeighbor clustering algorithm from {\tt scikit-learn} for $k=3$ was used to easily identify and compute the projected angular distances between the nearest stars for each of the 2,324 exoplanet systems (i.e. each exoplanet and its two nearest neighbors). Even using the ``brute force'' algorithm, this clustering takes less than 1 second to compute for such small samples of objects. Since many of the exoplanets in the sample reside in multi-planet systems, the NearestNeighbor algorithm will produce clusters with multiple object separations of exactly zero. I therefore further restricted my sample to clusters that contain no multi-planet systems, leaving a total of 1,217 clusters constructed of three single transiting systems. Note also that due to the uniform but stochastic sampling of exoplanet systems across the field (see \S\ref{sec:2pt}), some clusters are partially or fully redundant (i.e. the same three stars can be selected in differing orders in up to three separate clusters). I do not eliminate this redundancy in this demonstration.

\begin{figure}[!t]
\centering
\includegraphics[width=3in]{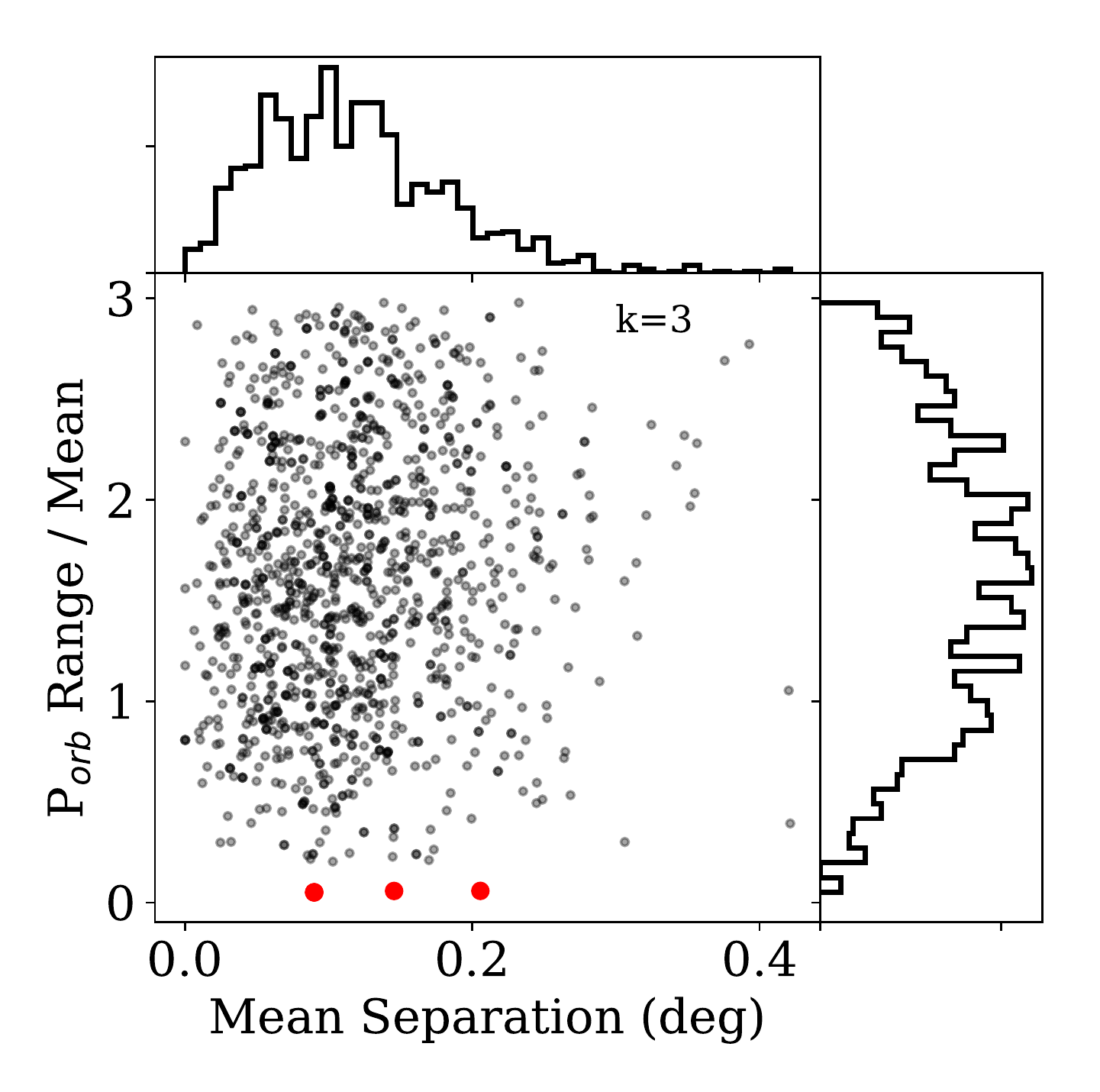}
\caption{Fractional range of orbital periods versus the mean angular separation for KNN clusters in (RA, Dec) space of transiting exoplanet systems in \Kepler. This sample only includes systems with only a single known transiting exoplanet. These KNN clusters have typical spatial separations of $\sim$0.1 deg. For $K$=3, three unique clusters have a spread in their orbital periods below 6\% (red points). However this is not a significant outlier in the orbital period similarity, and these clusters are not unusually close together in their spatial separation.}
\label{fig:knn}
\end{figure}

For each $k=3$ star cluster I compute two metrics that should be considered when searching for  spatially distributed technosignature beacons, shown in Figure \ref{fig:knn}. First I simply calculate the mean cartesian distance of the three exoplanet systems about the ``center'' or average spatial position in 2D space (RA, Dec). This enables us to potentially identify clusters of transiting systems that are unusually close together in projected separation. Second I compute the similarity of the orbital periods for the three exoplanets in each cluster. This similarity metric here is defined as the range (i.e. max period -- min period) divided by the mean orbital period of the three stars.
This period metric is easy to interpret, allowing us to identify clusters of transiting exoplanets whose orbital periods deviate by less than a given percentage.

I find no anonymously self-similar $k=3$ clusters of exoplanets in the \Kepler sample considered here. The cluster with the most similar period distribution consists of Kepler-1295, Kepler-1612, and Kepler-612. This triplet of systems have orbital periods within 5.1\% of each other (mean period of $P=3.82$ days), only $2.4\sigma$ from the average of the period range distribution in Figure \ref{fig:knn}.
This cluster also has a mean separation of 0.09 deg, only slightly closer than the typical cluster separation of 0.11 deg.

\section{Discussion}
\label{sec:discussion}

In this paper I have presented the great potential for using algorithms to search for SETI or ``technosignature'' signals in modern spatio--temporal surveys, such as TESS, ZTF, or LSST. Though these surveys are primarily in optical wavelengths as opposed to traditional SETI at radio frequencies, there are many promising algorithmic approaches that may be considered. Further, these searches can ``piggyback'' on datasets and computational resources being developed for other science goals.
This SETI approach also connects with the rise of data science and statistical methods in astronomy to characterize, and find outliers in, large datasets. Interestingly, the search for technosignatures is possibly one of only a few areas where ``big data'' survey astronomy intersects with the growing astrobiology discipline. As \citet{berea2019} note, expertise in many complementary domains of data analysis (e.g. cybersecurity) may also be useful for advancing survey SETI.

To illustrate the sorts of SETI algorithms that could be easily constructed based on existing data analysis and machine learning libraries, I have outlined a few simple approaches to search for beacons based on the distribution of exoplanets in the \Kepler field.
The most detailed example in \S\ref{sec:sim} searches for clusters of transiting exoplanet systems with unusually similar properties.
One obvious simplification of this approach as outlined is the need to pre-define the number of stars in each cluster, here assumed to be $k=3$. However, it is trivial in practice to adjust this number, or scan over many possible values. Indeed, I explored $k=2$ through $k=8$ clusters with this data, and found no substantial outliers in either the orbital period similarity or spatial separation.

While I've explored self-similarity between spatially proximate exoplanet systems as the search metric here, one could instead pick densities of the {\it most unlikely} transiting systems. Since long-period eclipses are geometrically the most rare to detect, a spatial cluster of long-period exoplanets should immediately stand out as being very unlikely, even if the period similarity and average separation is not exceptional. Adding such a long-period or unlikely-period cluster-finding algorithm to a future survey SETI toolkit is straightforward. However, these systems are also the most time-intensive to detect for an observer, and so a more efficient beacon for both construction and detection would consist of short-period systems. 

As described in the introduction, sky surveys enable a wide range of algorithmic approaches to use for SETI. Real-time alert streams from surveys like ZTF and soon LSST \citep[e.g.][]{schwamb2019} make rapid identification and follow-up of unusual variability straightforward to implement. Other approaches could include monitoring for over-densities in the spatial distribution of alerts, especially within the Earth Transit Zone \citep{heller2016}, searching for slowly appearing or disappearing stars on decades timescales \citep{villarroel2016}, or looking for coordinated variability with e.g. novae along the ``SETI ellipsoid'' over wide areas of the sky \citep{lemarchand1994}. Unnatural motions of objects could also be searched for in sky surveys. This might be akin to searching for interstellar asteroids \citep[e.g.][]{mamajek2017} whose trajectories suddenly change. In future work we will explore a framework for searching the ZTF alert stream for unusual or ``impossible'' sequences of events in light curves.

Finally, while the opportunity for conducing SETI research using large surveys has been noted for nearly two decades \citep[e.g.][]{djorgovski2000}, what has been lacking is the development of general-use algorithms or software packages designed for such searches. This is an area in clear need of further exploration, and an opportunity for contribution to SETI by many software-minded researchers from all disciplines.

\software{Python, IPython \citep{ipython}, NumPy \citep{numpy}, Matplotlib \citep{matplotlib}, SciPy \citep{scipy}, Pandas \citep{pandas}, Astropy \citep{astropy}}

\acknowledgments

J.R.A.D. thanks Nicholas M. Law and Jason T. Wright for their ongoing conversations in developing this idea, Andrew P. V. Siemion and the Berkeley SETI Research Center team for hosting me for an enlightening and motivating visit, the Research Corporation for Scientific Advancement for hosting the 2018 TDA Scialog meeting, and Daniela Huppenkothen for comments that greatly improved this manuscript.

J.R.A.D. acknowledges support from the DIRAC Institute in the Department of Astronomy at the University of Washington. The DIRAC Institute is supported through generous gifts from the Charles and Lisa Simonyi Fund for Arts and Sciences, and the Washington Research Foundation.

This paper includes data collected by the Kepler mission. Funding for the Kepler mission is provided by the NASA Science Mission directorate.

This research has made use of the NASA Exoplanet Archive, which is operated by the California Institute of Technology, under contract with the National Aeronautics and Space Administration under the Exoplanet Exploration Program.

This work made use of the gaia-kepler.fun crossmatch database created by Megan Bedell.


\end{document}